\documentstyle[twocolumn,prc,aps,epsfig]{revtex}
\input epsf
\newcommand{\be}{\begin{eqnarray}}
\newcommand{\ee}{\end{eqnarray}}
\def\lsim{\mathrel{\rlap{\lower3pt\hbox{\hskip1pt$\sim$}}
     \raise1pt\hbox{$<$}}} %less than or approx. symbol
\def\gsim{\mathrel{\rlap{\lower3pt\hbox{\hskip1pt$\sim$}}
     \raise1pt\hbox{$>$}}} %greater than or approx. symbol

\begin{document}

\twocolumn[\hsize\textwidth\columnwidth\hsize\csname 
@twocolumnfalse\endcsname

\title{Classical Strongly Coupled QGP I : \\
The Model and  Molecular Dynamics Simulations }

\author{Boris A. Gelman, Edward V. Shuryak and Ismail Zahed}
\address {Department of Physics and Astronomy\\
State University of New York, Stony Brook, NY 11794-3800}

%\date{\today}
\maketitle
\begin{abstract}
We propose a  model for the description of strongly interacting
quarks and gluon quasiparticles at $T=(1-3)T_c$, as a classical
and nonrelativistic colored Coulomb gas. The sign and strength
of the inter-particle interactions are fixed by the scalar product
of their classical {\it color vectors} subject to Wong's equations.
The model displays a number of phases as the Coulomb coupling is
increased ranging from a gas, to a liquid, to a crystal with
antiferromagnetic-like color ordering. We analyze the model
using Molecular Dynamics (MD) simulations and discuss the 
density-density correlator in real time. We extract pertinent 
decorrelation times, diffusion and viscosity constants for all 
phases. The classical results when extrapolated to the sQGP
suggest that the phase is liquid-like, with a diffusion constant 
$D\approx 0.1/T$ and a bulk viscosity to entropy density ratio 
$\eta/s\approx 1/3$.
\end{abstract}
\vspace{0.1in}
]
\begin{narrowtext}
\newpage

\section{Introduction}

The Quark-gluon plasma (QGP) is a high temperature
phase of QCD. Here, the word {\it plasma} is used 
in the same sense as in electrodynamical  plasma,
that is a phase with free  color charges that 
are screened~\cite{Shu_QGP} rather than confined
by the medium. Lattice simulations have shown, that 
the QGP exists above the critical temperature $T>T_c\approx 170\, MeV$
of the QCD phase transition,
in the deconfined and chirally symmetric phase.

Asymptotic freedom of Non-Abelian gauge theories, a property of QCD,
insures that for high enough temperature $T\gg \Lambda_{QCD}$ this phase 
becomes {\em weakly} coupled (wQGP), with most of the particle 
interactions characterized by a small coupling $\alpha_s(p\approx T)\ll 1$. 
In this regime, the wQGP is a  near-ideal gas
of its fundamental constituents, quarks and gluons.   
Perturbative methods, such as hard thermal loops~\cite{BP} and other
resummations techniques,  show how quarks and gluons are dressed and
become quasiparticles with $T$-dependent dispersion curves and widths.

The QGP is experimentally studied using heavy ion collisions, such 
as dedicated experiments carried at the CERN SPS during the
 previous decade, 
and now at the BNL RHIC collider. RHIC can reach temperatures 
of about $2T_c$. The success of the hydrodynamical description for
the observed collective flows at RHIC~\cite{hydro} has shown that all
dissipative lengths are very short. The produced matter
at RHIC {\it cannot} be a weakly coupled gas but a rather good liquid
\cite{Shu_liquid}. Recently, two of us~\cite{SZ_newqgp} have suggested 
that the interaction of quasiparticles in the relevant temperature
range at RHIC is strong enough to
%E account for the short dissipative lengths,
 generate multiple marginal colored Coulomb 
bound states. Some of those states (charmonium) are observed
in current lattice~\cite{charmonium} simulations till about $2.5 T_c$.
The existence of these states, especially the colored ones, is
still debated and warrant further numerical and independent checks 
on the lattice.

The effective potential energy of two static colored charges
separated by a distance $R$, $U(R)$ can also be deduced
from lattice simulations. Close to the critical temperature,
the separation energy $\Delta U$
(the potential at infinity minus its value at some typical
distance 0.3 \, fm) is $\Delta U\approx 4-1$ GeV in the
temperature range $T=(1-1.2)\, T_c=0.17-0.21$ GeV. The ensuing 
Boltzmann penalty $e^{-u}$ with $u= \Delta U/T\approx 20-5$
is mighty, indicating the dominance of potential over kinetic
energy. This regime is now called a {\em strongly} coupled QGP (sQGP).
Its structure and consequently its {\it transport properties} 
are radically different from a wQGP to which it is expected to fold at 
very high temperature.

In the traditional context  of an electromagnetic plasma,
the term ``strongly coupled'' plasma has a similar meaning. 
Ionic or dusty plasmas have charged ions with  large masses, 
and thus are essentially classical. The standard dimensionless parameter
characterizing the strength of the interparticle interaction in a
classical plasma is $\Gamma$ the ratio of potential to kinetic
energy

\be 
\label{eqn_Gamma} 
\Gamma= \frac{(Ze)^2}{a_{WS}T}
\ee
where $Ze,a_{WS},T$ are respectively the ion charge, the Wigner-Seitz radius
$a_{WS}=(3/4\pi n)^{1/3}$  and the temperature~\footnote{We do not use $k_B$ 
and thus measure temperature in energy units.}. $\Gamma$ is convenient 
to use because it only involves the {\it input} parameters, such as the 
temperature and density. However, one should keep in mind that the real 
interaction parameter is the {\it output} dimensional parameter $u$,
 
\be \label{eqn_u} 
u={U \over T}
\ee  
where $U=<V>$ is the average interaction of a particle with all its
neighbors. This is the ratio that enters the effective Boltzmann 
exponent, and defines all correlation functions. 

Since $u$ is proportional to $\Gamma$, one usually
defines the weakly coupled regime for $\Gamma\ll 1$ and
the strongly coupled regime in the opposite limit $\Gamma\gg 1$. 
Extensive studies of the one-component plasma (OCP) in electrodynamics,
using both MD and analytical methods over the past decades,
have revealed the following regimes:
{\bf i.} a gas regime for $\Gamma<1$; {\bf ii.} a
liquid regime for $\Gamma\approx  10$;  {\bf iii.} a glass regime
for $\Gamma\approx 100$; {\bf iv.} a solid regime for $\Gamma > 300$. 
For a review  see e.g.~\cite{Ichi}.

Another physical system closer to what we will discuss in this work is a
classical two-component plasma (TCP) with both positive and negative charges.
Examples are  molten or frozen salts with ions of comparable masses, as in
the sQGP. Hydrogen plasmas are better studied, but the underlying charges
carry very different masses, so the ensuing results bear no insight to the
sQGP were the masses are likely comparable.

 Quantum effects  will be schematically 
incorporated in the form of a {\it localization
energy}, providing a repulsive core for the 2-particle interaction
irrespective of the charge. In the future we plan to study more
quantum corrections to classical MD, in particular
the role of the exclusion principle, and 
the relation between classically {\it correlated charges}
and quantum bound states.

 We are certainly aware of many cases of quantum plasmas.
An quantum electron plasma 
in metals is quantum and  thoroughly studied, but
 repulsive nature of the electron-electron interaction 
excludes the formation of bound states and is very different from
QGP we are interested in. Only weak induced electron-electron
interaction leads Cooper pairing in superconductors.  Excitons
in semiconductors, in a plasma of particles and holes,
 are even closer to our problem.
   Recent progress in trapped strongly coupled 
fermionic atoms have further 
elucidated how the transition to Bose-condensed
pairs takes place. However, in all these cases one needs novel
tools to study quantum effects dynamically: the existing ones,
such as e.g. restricted paths Monte Carlo, 
use Euclidean time correlators which (like those from lattice QCD)
are good for thermodynamical observables but next-to-impossible to
use for transport properties.
The MD approach, whatever crude
it can be, is an invaluable classical tool
 which directly provides real-time
correlators. Furthermore,
pertinent quantum corrections can be added later, as it is done in
some atomic systems.

The model we propose in this work combines features of strongly 
coupled Abelian TCP, with non-Abelian features proper to QCD in the
form of classical color vectors for the underlying charges.
The main interparticle interaction is proportional to the dot-product 
of the color vectors, which of course can be attractive, repulsive or
even null if the color vectors of the particles are orthogonal to each other. 
Dynamics of color vectors as well as particle coordinates 
are described by classical equations of motion (EoM), and thus can be 
studied by MD simulations, which are less involved than the corresponding
coupled hierarchy of quantum Green's functions.
Below, we will argue for which values of the parameters this model
can be useful for understanding the sQGP. Before we will do so,
the model will be defined and studied in its own right.

\section{Classical Quark-Gluon Plasma, cQGP}

\subsection{The model}

The model is based on the following main assumptions:\\
{\bf i.} The particles are heavy enough to move 
nonrelativistically, with masses $M>>T$; \\
{\bf ii.} The inter-particle interaction is dominated by 
colored electric (Coulomb) interactions, with all magnetic 
effects (like e.g. spin forces) ignored;\\ 
{\bf iii.} The color representations are large, so that
color operators $t^a$ can be represented by their
average, classical color vectors.\\

The parameters of the model include the particle mass
$M$ and their density $n$.  The main interaction potential
is proportional to the dot-product of the unit color vectors
times the standard Coulomb interaction strength. The notations 
we use parallel the conventions in EM plasmas and in QCD.
The relation between the two is defined as

\be 
(Z_\alpha e)^2 = C_\alpha {g^2 \over 4\pi} 
\ee
where  $Z^2_\alpha$ in EM is $C_\alpha$ in QCD,
the Casimir operator eigenvalue for 
$\bar q, q, g$, while $e^2=g^2/4\pi$~\footnote{Note
that the standard notations used in EM and QCD differ by
$4\pi$ in the Lagrangian and all subsequent formulae.}.
When needed, we will ``input'' the Debye screening
mass'' $M_D$, if the classical screening of plasma charges 
derived dynamically from MD would not be an accurate representation
of the effective forces between quasiparticles in the sQGP.

In QCD, all the effective parameters used here are functions 
of the temperature $T$ and the baryon chemical potential $\mu$.
or of  a point at the QCD phase diagram. In a heavy ion
 collision, those are defined by the collision energy and
centrality of the collision. Furthermore, for each volume element
the cooling of the matter during its expansion is well approximated by
the adiabatic (fixed entropy/baryon densities)
 path on the phase diagram, related $T$ and $\mu$.

As a first approximation, baryonic charge play little role at RHIC
and thus one can set $\mu=0$ and think of only one parameter, 
 the temperature $T$, defining matter properties.
Therefore, when we attempt to map  the MD
results for the cQGP into sQGP,
 all the parameters of the model mentioned above should
be converted to temperature, as we will detail. In way, the 
classical model we are about to discuss in an unrestricted
n-dimensional parameter space, will be useful for the sQGP
through a 1-dimensional slicing of the n-dimensional space.
Probing the n-dimensional parameter space for the model is
useful theoretically. Much like EM plasmas, the dimensionless
parameter (\ref{eqn_Gamma}) will be key in characterizing the 
transition from a weakly coupled regime with $\Gamma\ll 1$,
to a strongly coupled regime $\Gamma\gg 1$.

At very low temperatures $T$ or very large $\Gamma$,   
any classical system will freeze. Therefore, the low
temperature behavior is dominated by the lowest ordered
state.
Here, we recall that in general, the Abelian 2-component 
plasma freezes to an ionic crystal much like ordinary salt $Na Cl$.
The non-Abelian plasma under consideration freezes also to 
{\em cubic crystal} with {\em ferromagnetic} (alternating) order
of the classical color vectors. A direction in classical color space is 
selected  randomly, through spontaneous symmetry breaking of the global 
color group, with alternating  directions of the color vectors along 
the crystal axes.

\subsection{The equations of motion}

The specific model Hamiltonian is

\be 
H=\sum_{\alpha\,i}\frac{p_{\alpha\,i}^2}{2m_\alpha} + V_C +V_{\rm core} 
\label{HAMIL}
\ee
with the standard kinetic energy, the colored Coulomb interaction $V_C$
and the repulsive core $V_{\rm core}$ respectively. The colored
Coulomb part is

\be
V_C=
\sum_{\alpha\,i\neq\beta\,i}% V_{\alpha,i,\beta,j}
\frac{Q_{\alpha\,i}^a\,Q_{\beta\,j}^a}{|\vec{x}_{\alpha\,i}
-\vec{x}_{\beta\,j}|}
\label{HCB}
\ee
The sums are over the species $\alpha=q,\overline{q},g$ and their
respective numbers $N_\alpha$. We will specify $V_{\rm core}$ below.

Thus our phase space coordinates are the
position ($x_\alpha$), momentum ($p_\alpha$) and color ($Q_\alpha$).
The latter rotates under a gauge transformation

\be
Q^a\rightarrow D^{ab}(\Lambda)\,Q^b
\label{rot}
\ee
showing that (\ref{HCB}) is gauge invariant. Only the Coulomb-like
interaction was retained in (\ref{HCB}) since the magnetically
induced interactions are subleading non-relativistically. This
is the case for all spin and {\rm local} many-body forces. for 
instance the {\rm local} 3-body gluon-induced interaction

\be
\sum_{\alpha\,i\neq\beta\,j\neq\gamma\,k}
Q_{\alpha_i}^a\,Q_{\beta_j}^b\,Q_{\gamma_k}^c\,
\left(\,f^{abc}\,{\bf F}+d^{abc}\,{\bf D}\right)
\label{three}
\ee
is subleading ${\bf F}\approx {\bf D}\approx \nabla/m$. Non-local
many-body interactions as induced by the 2-body Coulomb and core
interactions are important and will be resummed to {\rm all orders}
using MD.

The EoM for the phase space coordinates follow from the usual 
Poisson brackets. For the standard coordinates they are

\be
\{  x_{\alpha\,i}^m, p_{\beta\,j}^n \}=\delta^{mn} 
\delta_{\alpha\beta}\delta_{ij}  
\ee
For the color coordinates they are 

\be 
\{ Q_{\alpha\,i}^a, Q_{\beta\,j}^b\}= f^{abc}\,Q_{\alpha\,i}^c
\label{can1}
\ee
the classical analogue of the SU(N$_c$) color commutators,
with $ f^{abc}$ the known structure constants of the color group.
The classical color vectors are all adjoint vectors with
$a=1...(N_c^2-1)$. For simplicity only the non-Abelian group SU(2)
will be considered, with 3d color vectors. The difference between quarks
and gluons follow from their respective Casimir assignments.

Although the brackets (\ref{can1}) do not look like the usual
canonical relations between coordinates and momenta, they actually
are. Indeed, a more standard  phase space description requires the
use of the Darboux parameterization~\cite{darboux}

\be
Q=\left( \Phi^A,\Pi^A\right)
\label{can2}
\ee
with $A=1,..,N_c(N_c-1)/2$ now satisfying standard canonical commutation 
relations

\be
\{\Phi^A,\Pi^B\}=\delta^{AB}
\label{can3}
\ee
the $\Phi's$ are angular coordinates and the $\Pi's$ are angular momenta.
The  $\Pi's$ are identified with the fixed Casimirs of $SU(N_c)$ which are
conserved by the equations of motion.  For example, for
$N_c=2$, the 3 components of the color vector conserve one Casimir
operator $C_2$ (the color vector's length). Without loss of generality,
the Darboux coordinates maybe chosen as

\be
(Q_1\pm i Q_2, Q_3)=(\sqrt{J^2-\Pi^2}e^{\pm i\Phi},\Pi)
\ee
with $J^2/3=C_2$. For $N_c=3$ there the color vector is 8-dimensional, 
with 2 fixed Casimirs $(Q^aQ^a)$ and $(d^{abc} Q^aQ^bQ^c)$, and 6 
variables. In this case, the Darboux set is more involved, but can
be parametrized with 3 angles and  3 conjugate momenta. 
We will not be working with  Darboux-based EoM below.

The pertinent equations of motion for our classical Coulomb gas
resulting from (\ref{HAMIL}) and the Poisson brackets are

\be
&&\dot{x}_{\alpha\,i}^n=\{H, x_{\alpha\,i}^n\}=
\frac{p_{\alpha\,i}^n}{m_\alpha}\nonumber\\
&&\dot{p}_{\alpha\,i}^n=\{H, p_{\alpha\,i}^n\}=
g\,E^{an}_{\alpha\,i}\,Q_{\alpha\,i}^a\nonumber\\
&&\dot{Q}_{\alpha\,i}^a=\{H, Q_{\alpha\,i}^a\}=
g\,f^{abc}\,Q_{\alpha\,i}^b\,A_{\alpha\,i\,0}^c
\label{eom}
\ee
with

\be
%E_{\alpha\,i}^a=-\vec\nabla_{\alpha\,i} Q_{\alpha\,i}^a V(x_{ij})
\vec{E}_{\alpha\,i}^a=&&-
\vec\nabla_{\alpha\,i}\,A_{\alpha\,i\,0}^a
\nonumber\\=&&
-\vec\nabla_{\alpha\,i}\sum_{\beta\,j\neq \alpha\,i}
\frac{g\,Q_{\beta\,j}^a}{|\vec x_{\alpha\,i}-\vec x_{\beta\,j}|}
\ee
The last of the three equation of motion  in (\ref{eom}) is
also known as Wong's equation~\cite{wong}.  Due to the antisymmetric
nature of the structure constant $f^{abc}$, it has the form of rotation 
of the color vector conserving its length (classical precession). 
For SU(2) this is the only Casimir, while for higher color representations
more Casimirs are involved. We expect the resulting EoM to generate a 
ballet of classical precession conserving all these Casimirs.

\section{Numerical studies using Molecular Dynamics}

In this paper we will outline the basic ingredients 
involved in the MD analysis of (\ref{eom}). We first
discuss what the pertinent scales are and how they are
set. Second, we briefly comment on the numerical 
analysis used to deal with long-range Coulomb forces.
 
\subsection{Potential and units}

The MD simulation is basically a numerical solution
of EoM  starting from some initial conditions.
The method was developed in the early days of
computers in the 1950's, and applied extensively 
for the studies of simple liquids and plasmas.
It is the most straightforward way to access 
transport properties such as the diffusion coefficient
$D$, the bulk viscosity $\eta$, the heat conductivity
$\lambda$, as well as pertinent decorrelation functions
and times.

In standard 2-component plasma physics MD simulations, 
the interparticle potential is chosen as

\be 
V(r)=
V_{\rm core}+V_C=\left(\frac{e^2}{\lambda}\right)\left[{1\over n} 
\left({\lambda \over r}\right)^n +{Q_i  Q_j\over r} \right]\ee
with $r=|\vec x_i-\vec x_j|$ and $Q_i,Q_j=\pm 1$ are respective
charges. The repulsive core potential $V_{core}$ ensures
overall matter stability with $n$ some large parametric number.
For atomic systems $n=9$. For molten salts the core follows
from the intrinsic repulsion of the atomic electrons.

Since this is the first paper in a series on this model, and 
as we are mostly interested in understanding the role played by 
the non-Abelian color variables, we have carried the numerical 
MD analysis for both the Abelian and non-Abelian charges. Both
analysis were carried in parallel to understand the differences
between discrete $\pm $ charged plasmas and our classically
colored SU(2) plasma with $Q^a$ charges. Therefore, the same 
$V_{\rm core}$ was kept in the non-Abelian simulation.

The microscopic motivation in QCD for $V_{\rm core}$ stems
from the short range part of the inter-particle interactions.
Needless to say that this is a hard question, and not much is
known about it from first principles whether in the intermediate 
coupling region ($\alpha_s\approx 1/2$) or the very strong coupling
region ($\alpha_s\approx 1$ or $\Gamma\gg 1$). 

  One reason for an effective
repulsion at short distance is the so called {\rm quantum
localization} potential

\be  
V_{\rm loc} \approx {\hbar^2 \over m r^2}\,\,. 
\ee
Such potentials are usually used in classical studies/
simulations of light
atoms, e.g. solid He. Note that this term has a lower power $n=2$ 
rather than $n=9$ used in $V_{\rm core}$. This not a much importance
as any $n>1$ stabilizes the Coulomb attraction at short distances.
The net effect of the quantum localization is the introduction of
a dimensionful parameter, i.e. $hbar$ in natural classical units.

{\bf i. Length unit :}
At any $n$, the minimum of the potential 
is at $r=\lambda$. So $\lambda$ sets the basic length scale, and 
from now on all distances are measured in units of $\lambda$. 
For example, a close cubic-packing density  with $n_{cp}=1/\lambda^{3}$ 
will be written as $n_{cp}=1$.
{\bf ii. Time unit :}
The unit of time $\tau_0$ will be set by the natural frequency
of a screened Coulomb system, its plasma frequency

\be 
\tau_0=\omega_p^{-1}= ({m\over 4\pi n e^2})^{1/2}
\ee
{\bf iii. Mass unit :}
This is naturally given by the particle mass $m$.
Thus all important time correlators have a characteristic length
of order 1 in $\tau_0$. In contrast, all MD runs are typically few 
hundreds or thousands $\tau_0$ to insure statistical equilibrium.

This  fixes the three basic units used in our analysis. For
instance, the kinetic energy is measured in obvious units of 
$m\lambda^2/\tau_0^2$. The  strength of the Coulomb potential, 
if measured in such units, is

\be 
({e^2\over \lambda})/({m\lambda^2\over\tau_0^2})=
{1\over 4\pi}{1\over n\lambda^3} 
\ee
so that it is defined directly by the particle density.
The temperature, in the same units, can be defined via averaged
(dimensionless) velocities as $(3/2)T=<v^2/2>$.  Finally
the main parameter $u$ in (\ref{eqn_u}) can be obtained from
simulation,
as  the ratio of the $measured$ total mean potential to the
total  kinetic energy.

\subsection{Technical details}

There are readily available numerical packages for solving
the set coupled set of differential equations of the type
(\ref{eom}). In our case, their accuracy is tested through 
energy conservation and color vectors remaining on the unit
sphere. We have used the package from the CERN Library in
double precision and tested its accuracy for both energy and
color conservation.

The many particle problem in a box requires the introduction
of boundary conditions. We chose periodic boundary conditions.
When a particle crosses one side, it reappears on the opposite
side. We can visualize space as filled with many mirror cubes,
piled next to each other. When one particle crosses out of the
central cube, its mirror image enters from the opposite cube. 
The inter-particle potential in this many-cube world is periodic
under cubic translation by $\vec{\bf L}$, i.e.

\be 
V_{ij}= 
\sum_{n_x,n_y,n_z=-\infty}^{+\infty} 
V(\vec x_i-\vec x_j + n\,{\vec{\bf L}})
\label{per}
\ee
So as a particle crosses the boundary, there is no change in
the force or the potential since (\ref{per}) is truly periodic.
In practice, the sum in (\ref{per}) is only carried over few
adjacent cubes or mirror images, causing an overall loss of 
periodicity. As a result, the forces at the faces of the cube
jump slightly thereby causing random kicks on the crossing 
particles. This gradually heats the system and should be 
controlled. The more long range the potential is,
 the more severe the heating.

Traditionally the Bethe polynomials for Ewald sums are used  
to overcome this heating phenomenon caused by the truncation.
We have found this procedure not sufficiently accurate for the
amount of complexity it introduces. Indeed, it leads to a factor
of 50 suppression of the electric field normal to the cubic 
interface. Current computers easily allow for keeping up to 3-4
mirror images on each side of a central cube, thereby reducing
the net electric field by  orders of magnitude. 

Another practical way to deal with this problem is
introduction of a very small friction into the EoM, which can 
compensate the heating effect caused by the random forces at
the cubic interface. Our typical runs consists of $4^4=64$ or 
$6^4=1296$ particles all of the same species.

\subsection{Structure factor}

The static and dynamic properties of a liquid are determined by 
various correlations functions. One of the most useful of these
functions is a density-density correlation function or structure
factor defined as,

\be
G(r,t)={1\over n} \left<\rho(\vec{x}, t) \, \rho(\vec 0 ,0) \right> \,,
\label{rho_rho}
\ee
with $r=|\vec x |$ and  $\rho(\vec{x},t)$ being
 the particle  density 
at the position $\vec{x}$ at time $t$,

\be
\rho(\vec{r},t)=\sum^{N}_{i=1} \delta(\vec{x}-\vec{x}_{i}(t)) \,,
\label{rho}
\ee
Here, $n=N/V$ is the particle density and the averaging
in~(\ref{Grt}) is carried over an equilibrium ensemble
~\footnote{In a given MD simulation the total energy and number 
of particles  of a system is kept constant. This corresponds to
using a microcanonical ensemble.}. 

Assuming translational invariance and isotropy, the classical 
density-density correlation function can be written as,
 
\be
G(r,t)={1\over N} \left<\sum^{N}_{i=1} \sum^{N}_{j=1} \,
\delta \left (\vec{x}+\vec{x}_{i}(0)-\vec{x}_{j}(t) \right) \right> \,,
\label{Grt}
\ee
with $N$ is the number of particles, $\vec{x}_{i}(t)$ is the position
of the ith-particle at time $t$. $G(r, t)$ characterizes the 
likelihood to find 2 particles a distance $r$ away from each other at 
time $t$.

\begin{figure}[ht]
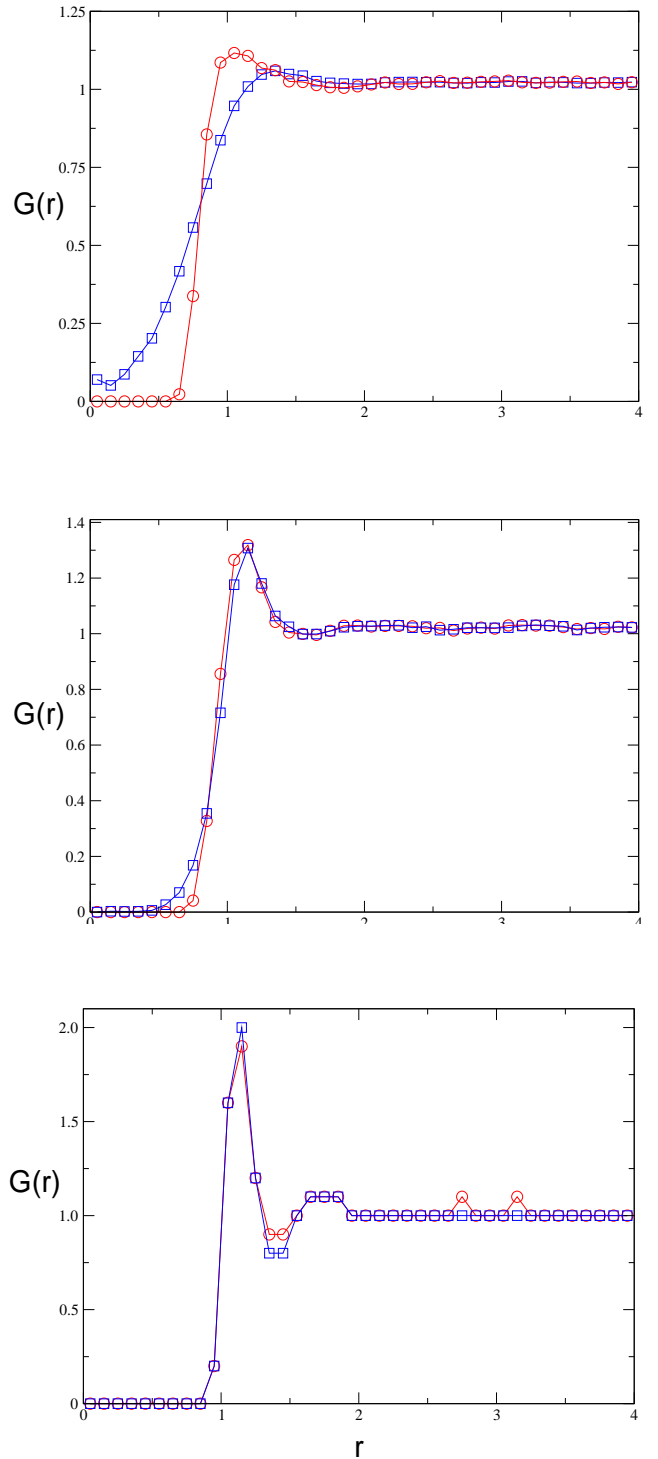

\begin{center}
\epsfig{figure=Gd_1.eps,height=6cm,width=8.4cm,angle=0}\\
\vspace{0.3in}
\epsfig{figure=Gd_2.eps,height=6cm,width=8.4cm,angle=0}\\
\vspace{0.2in}
\epsfig{figure=Gd_3.eps,height=6cm,width=8.4cm,angle=0}
\bigskip
\caption{$G_{d}$ correlation function for $\Gamma=0.83,31.3,131$,
respectively. Red circles correspond
to $t^{*}=0$, and blue squares correspond to $t^{*}=6$.}
\label{Gd1}
\end{center}
\end{figure}

The density-density correlation function~(\ref{Grt}) can 
be written as a sum of two parts referred to as the van Hove
or {\it self-correlation} $G_s$,  and the
{\it distinct-correlation} $G_d$,

\be
G(\vec{x},t)=G_{s}(\vec{x},t)+G_{d}(\vec{x},t)\,,
\label{GsGd}
\ee
where 
\be
G_{s}(\vec{x},t)={1\over N} \left<\sum^{N}_{i=1}  \,
\delta \left (\vec{x}+\vec{x}_{i}(0)-\vec{x}_{i}(t) \right) \right> \,,
\label{Gs}
\ee
and,
\be
G_{d}(\vec{x},t)={1\over N} \left<\sum^{N}_{i\neq j}  \,
\delta \left (\vec{x}+\vec{x}_{i}(0)-\vec{x}_{j}(t) \right) \right> \,.
\label{Gd}
\ee

%\vspace{0.2in}

%\begin{figure}[ht]
%\begin{center}
%\epsfig{figure=Gs_1.eps,height=6cm,width=8.4cm,angle=0}
%\bigskip
%\caption{$G_{s}$ correlation function for $G=0.83$ and $t^{*}=1$.}
%\label{Gs1}
%\end{center}
%\end{figure}  

%\begin{figure}[ht]
%\begin{center}
%\epsfig{figure=Gs_1a.eps,height=6cm,width=8.4cm,angle=0}
%\bigskip
%\caption{$G_{s}$ correlation function for $\Gamma=0.83$. Red circles correspond
%to $t^{*}=2$, green triangles correspond to $t^{*}=3$, and blue squares
%to $t^{*}=7$. }
%\label{Gs1a}
%\end{center}
%\end{figure} 

%\begin{figure}[ht]
%\begin{center}
%\epsfig{figure=Gs_2.eps,height=6cm,width=8.4cm,angle=0}
%\bigskip
%\caption{$G_{s}$ correlation function for $\Gamma=31.3$.
%Red circles correspond%
%to $t^{*}=2$, green triangles correspond to $t^{*}=3$, and blue squares
%to $t^{*}=7$.}
%\label{Gs2}
%\end{center}
%\end{figure}

%\begin{figure}[ht]
%\begin{center}
%\epsfig{figure=Gs_3.eps,height=6cm,width=8.4cm,angle=0}
%\bigskip
%\caption{$G_{s}$ correlation function for $\Gamma=131$. 
%Red circles correspond
%to $t^{*}=2$, green triangles correspond to $t^{*}=3$, and blue squares
%to $t^{*}=7$.}
%\label{Gs3}
%\end{center}
%\end{figure}  

We have measured a large selection of such function, as a typical
examples let us present here $G_d$ at three values of the coupling,
moderate , strong and very strong, $\Gamma=0.83,31.3,131$,
see Fig.\ref{Gd1}. One sees from those that in the first case
 a relatively weak 
correlation between the particles, at distance 1 corresponding to the
potential minimum, which relaxes rather quickly with time.
The correlation is more robust in the second ``liquid'' case,
and is very stable and is accompanied by extra peaks in a ``solid''
last case.

%\begin{figure}[ht]
%\begin{center}
%\epsfig{figure=Gd_2.eps,height=6cm,width=8.4cm,angle=0}
%\bigskip
%\caption{$G_{d}$ correlation function for $\Gamma=31.3$. Red circles correspond
%to $t^{*}=0$, and blue squares correspond to $t^{*}=6$.}
%\label{Gd2}
%\end{center}
%\end{figure}  

%\begin{figure}[ht]
%\begin{center}
%\epsfig{figure=Gd_3.eps,height=6cm,width=8.4cm,angle=0}
%\bigskip
%\caption{$G_{d}$ correlation function for $\Gamma=131$. Red circles correspond
%to $t^{*}=0$, and blue squares correspond to $t^{*}=6$.}
%\label{Gd3}
%\end{center}
%\end{figure}  

\subsection{Transport coefficients}

An important aspect of the strongly coupled plasmas is their
dramatic change in transport properties in comparison to weakly
coupled plasmas. The current MD simulations can be used to study 
the bulk transport properties at strong coupling. In particular,
the shear and bulk viscosities, diffusion constant, thermal 
conductivity, color conductivity etc. A simple way to obtain 
these transport coefficients is via Green-Kubo relations. These 
relations give the transport coefficients in terms of the integrals 
of the equilibrium time-dependent correlation functions. 

In the case of the self-diffusion the corresponding correlation
function is the velocity autocorrelation function,

\be
D(\tau)={1\over 3 N}\, \left<\sum^{N}_{i=1} \vec{v}_{i}(\tau)
\cdot  \vec{v}_{i}(0) \right> \, ,
\label{Dt}
\ee
where
$\vec{v}_{i}(\tau)$ is the velocity of a particle {\it i} at time 
$\tau$. The velocity autocorrelation functions are shown in
Fig.~\ref{Dt1} for 
$\Gamma= 0.83, 31.3, 131$ respectively. This refers to the 
non-ideal gas, liquid and crystal of the (one species) cQGP.

The diffusion constant is the integral of the velocity
autocorrelation function,

\be
D=\int_{0}^{\infty} D(\tau) \, d\tau  \,.
\label{D}
\ee

Fig.~(\ref{diffusion}) shows a log-log plot of the diffusion constant as 
a function of $\Gamma$ . The dependence of $D$ on $\Gamma$ is linear and
can be approximately described by a simple power

\be
D\approx {0.4 \over \Gamma^{4/5} }
\label{diffusion_fit}
\ee

%\vspace{0.3in}

\begin{figure}[ht]
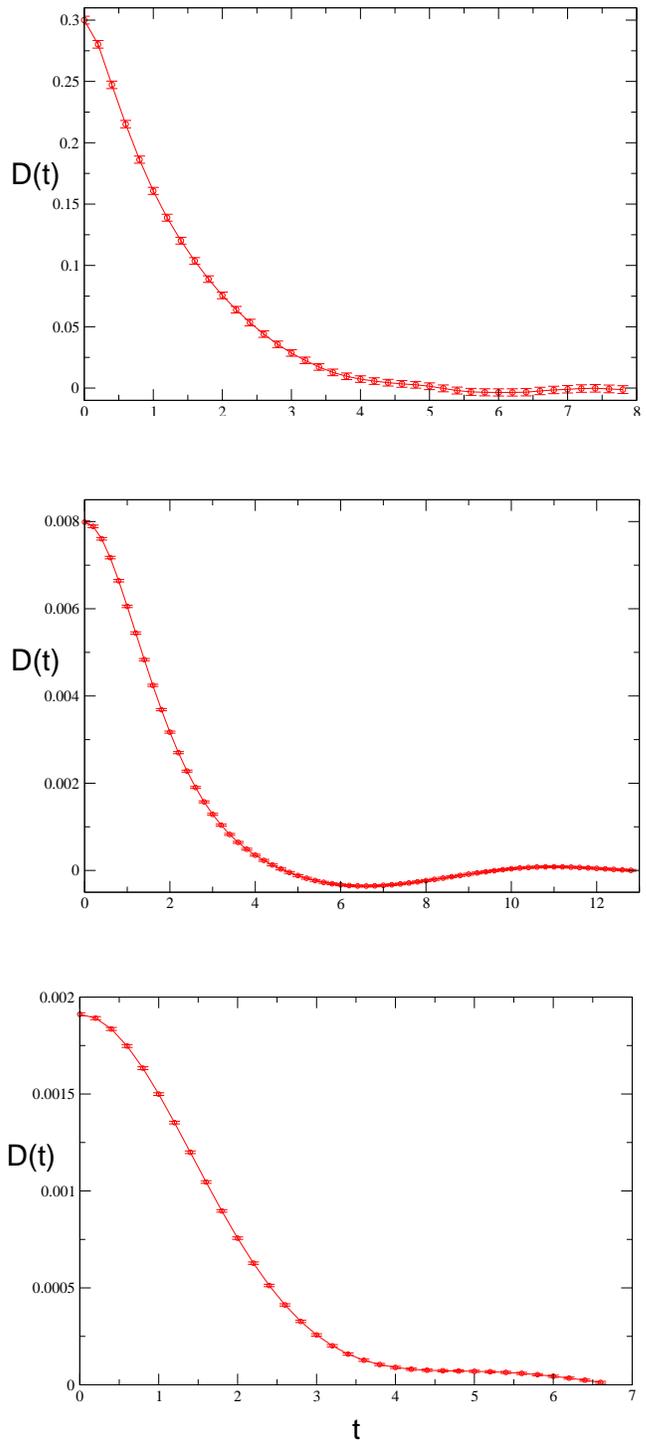

\begin{center}
\epsfig{figure=Dt_1.eps,height=6cm,width=8.4cm,angle=0}\\
\vspace{0.2in}
\epsfig{figure=Dt_2.eps,height=6cm,width=8.4cm,angle=0}\\
\vspace{0.2in}
\epsfig{figure=Dt_3.eps,height=6cm,width=8.4cm,angle=0}
\bigskip
\caption{Velocity autocorrelation function $D(\tau)$ for
$\Gamma=0.83, 31.3, 131$.}
\label{Dt1}
\end{center}
\end{figure} 

%\begin{figure}[ht]
%\begin{center}%

%\bigskip
%\caption{Velocity autocorrelation function $D(\tau)$ for
%$\Gamma=31.3$.}
%\label{Dt2}
%\end{center}
%\end{figure}%

%\begin{figure}[ht]
%\begin{center}

%\bigskip
%\caption{Velocity autocorrelation function $D(\tau)$ for
%$\Gamma=131$.}
%\label{Dt3}
%\end{center}
%\end{figure}

%\vspace{0.2in}

\begin{figure}[ht]
\begin{center}
\epsfig{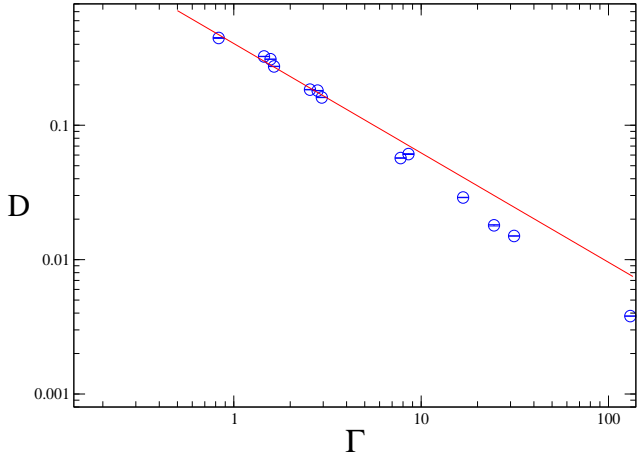}
\bigskip
\caption{The diffusion constant of a one species cQGP as a function of
the dimensionless coupling $\Gamma$.  Blue points are the MD simulations; the
red curve is the expression~(\ref{diffusion_fit}).}
\label{diffusion}
\end{center}
\end{figure}

\vspace{0.2in}

The viscosity coefficients are given in terms of
time autocorrelation function of the stress-energy tensor,

\be
\eta(\tau)={1\over 3\, T V}\, \left<\sum_{x<y} 
\sigma_{xy}(\tau) \, \sigma_{xy}(0)\right> \, ,
\label{etat}
\ee
where $\sum_{x<y}$ denotes a sum over the three pairs of distinct 
tensor components ($xy,yz$ and $zx$). The off-diagonal parts
stress-energy tensor are given by

\be
\sigma_{xy}= \sum_{i=1}^{N} m_{i} v_{ix} v_{iy} + {1\over 2} \,
\sum_{i\neq j} r_{ij,x} F_{ij,y}   \, ,
\label{sigmaxy}
\ee
and cyclically, with $\vec{r}_{ij}=\vec{x}_j - \vec{x}_i$ and $\vec{F}_{ij}$ is 
the force on particle {\it i} due to particle {\it j}.
The stress-tensor correlation functions are shown 
in Fig.~\ref{etat1} for $\Gamma=0.83, 31.3, 131$ respectively.

The Green-Kubo relation for the the coefficient of the shear viscosity is,

\be
\eta=\int_{0}^{\infty} \eta(\tau) \, d\tau  \,.
\label{eta}
\ee 
The coefficient of shear viscosity as a function of $\Gamma$ is
shown in fig.~\ref{viscosity}. For small $\Gamma$ the viscosity is 
large since the mean-free path is large in a gas-like phase. The
viscosity is minimum in the liquid phase, and rises slowly in the
crystal phase at large $\Gamma$.  The coefficient 
$\eta$ is approximately fitted by

%\vspace{0.2in}

\begin{figure}[ht]
\begin{center}
\epsfig{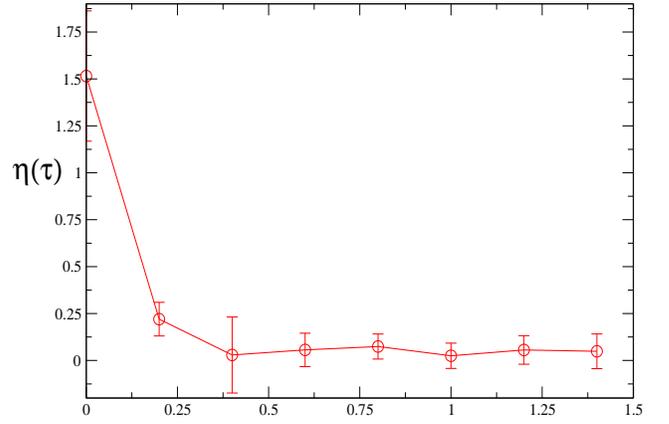}\\
\vspace{0.2in}
\epsfig{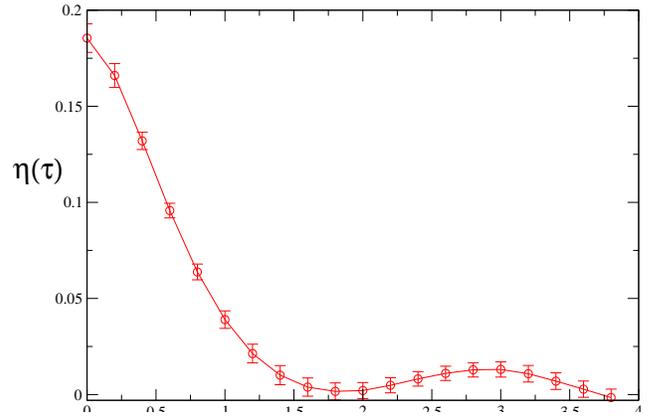}\\
\vspace{0.2in}
\epsfig{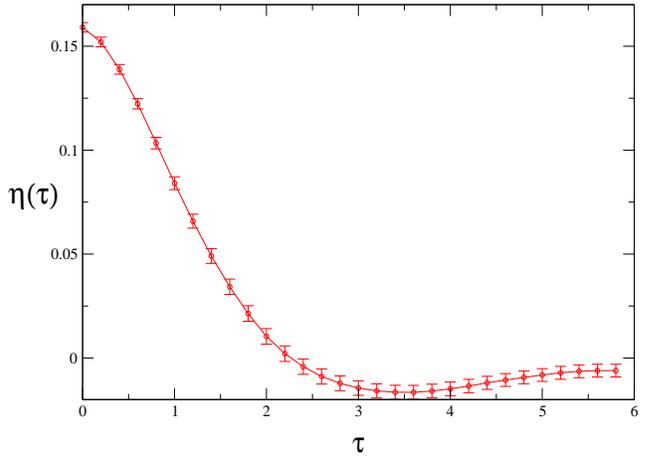}
\bigskip
\caption{Stress-tensor autocorrelation function $\eta(\tau)$ for
$\Gamma=0.83, 31.3, 131$.}
\label{etat1}
\end{center}
\end{figure}

%\begin{figure}[ht]
%\begin{center}

%\bigskip
%\caption{Stress-tensor autocorrelation function $\eta(\tau)$ for
%$\Gamma=31.3$.}
%\label{etat2}
%\end{center}
%\end{figure}

%\begin{figure}[ht]
%\begin{center}

%\bigskip
%\caption{Stress-tensor autocorrelation function $\eta(\tau)$ for
%$\Gamma=131$.}
%\label{etat3}
%\end{center}
%\end{figure}

\be
\eta \approx 0.001 \, \Gamma +\frac{0.242}{\Gamma^{0.3}}+
\frac{0.072}{\Gamma^{2}}\,.
\label{viscosity_fit}
\ee
Notice that the fit misses $\eta\approx 0.1$ at $\Gamma=10$ which is
the lowest viscosity measured in the current cQGP.

\vspace{0.3in}

\begin{figure}[ht]
\begin{center}
\epsfig{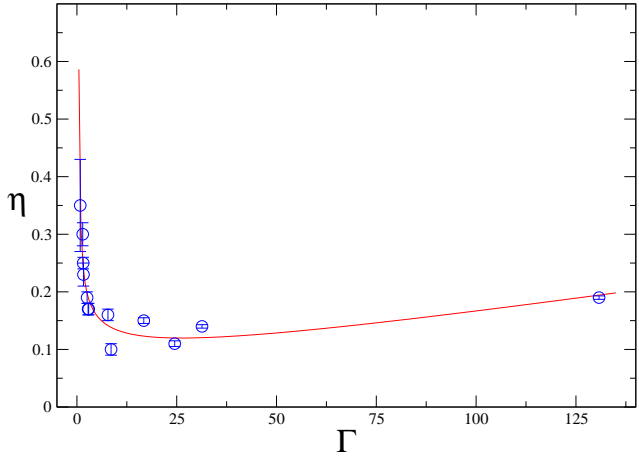}
\bigskip
\caption{Shear viscosity of one species cQGP as a function of
the dimensionless coupling $\Gamma$. Blue points are the MD simulations; the
red curve is the fit in eq.~(\ref{viscosity_fit}).}
\label{viscosity}
\end{center}
\end{figure}

The stress-tensor auto-correlation function can be used
to determine the viscous decorrelation time $\tau_{\eta}$,
defined as

\be
\tau_{\eta}= {\eta \over \eta(0)} \, ,
\label{taueta}
\ee
where $\eta(0)$ is the value of the stress-energy auto-correlation
function at time $t=0$. The values of  $\eta(0)$ and $\tau_{\eta}$
as a function of $\Gamma$ are shown in Fig.~\ref{visc0} and 
Fig.~\ref{taueta_fig}, respectively.

\vspace{0.3in}

\begin{figure}[ht]
\begin{center}
\epsfig{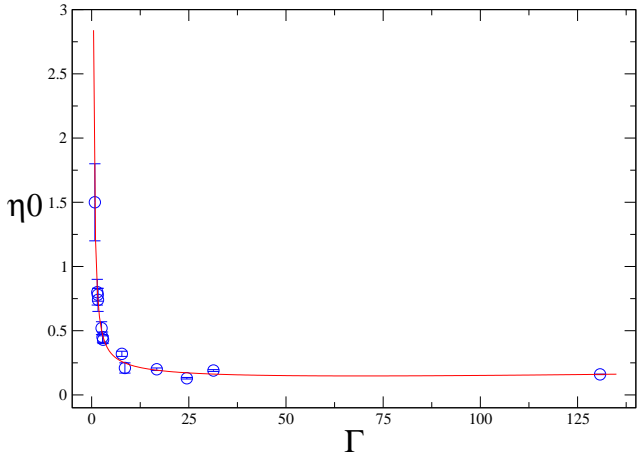}
\bigskip
\caption{Values of stress-energy auto-correlation at zero time 
of a one species cQGP as a function of
the dimensionless coupling $\Gamma$.  The blue points are the 
MD simulations; the red curve is the fit in eq.~(\ref{visc0_fit}).}
\label{visc0}
\end{center}
\end{figure} 

The functional dependence on $\Gamma$ of $\eta(0)$ is

\be
\eta (0) \approx 0.0005 \, \Gamma + {0.77  \over \Gamma^{1.57}}+
{0.44  \over \Gamma^{0.33}} \, .
\label{visc0_fit}
\ee
while that of the decorrelation time is

\be
\tau_{\eta} \approx 0.239 + 0.091 \, \sqrt{\Gamma} \, .
\label{tau_eta_fit}
\ee

\vspace{0.2in}

\begin{figure}[ht]
\begin{center}
\epsfig{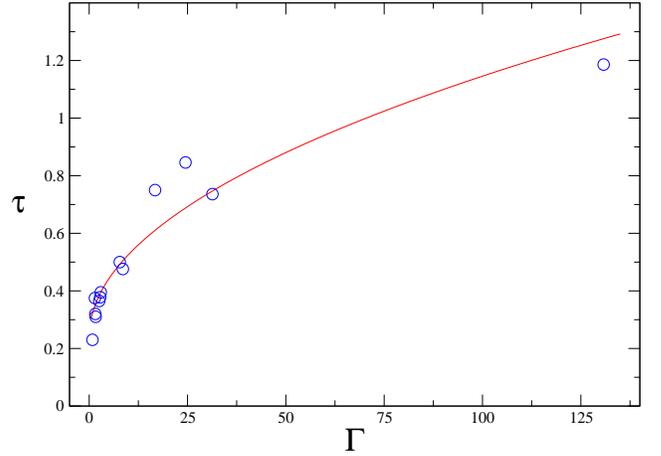}
\bigskip
\caption{The viscous decorrelation time for the stress-energy tensor time 
correlator as a function of the dimensionless coupling $\Gamma$.
The blue points are the MD simulations; the red curve is the fit in
eq.~(\ref{tau_eta_fit}).}
\label{taueta_fig}
\end{center}
\end{figure}

\subsection{Potential energy}

Another useful observable that can be determined from MD simulations
is the total potential energy $U$ normalized to $T$. This ratio
as a function of $\Gamma$ is shown in Fig.~\ref{uex}. The parametric 
dependence of this ratio on $\Gamma$ is given by,

\be
{U\over N\, T} \approx -4.9-2\,\Gamma +3.2\, \Gamma^{1/4}
+{2.2\over \Gamma^{1/4}} \,.
\label{uex_fit}
\ee 
At large $\Gamma$ the crystal phases sets in, and the potential energy
asymptotes $U/NT\approx -2\Gamma$ which is a measure of the Madelung 
constant. A full theoretical analysis of (\ref{uex_fit}) will be given in
the sequel II.

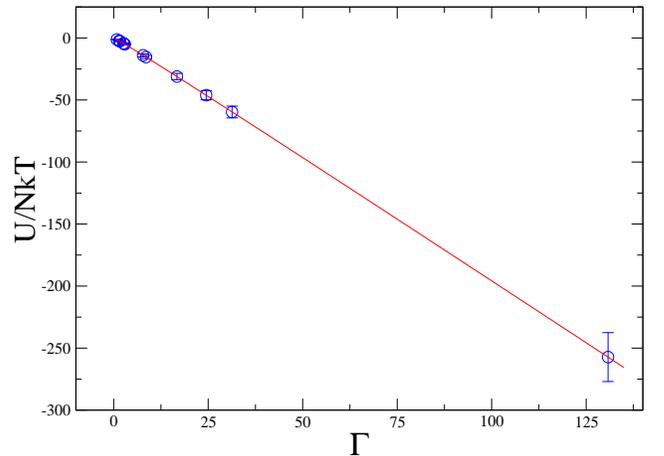
\vspace{0.3in}

\begin{figure}[ht]
\begin{center}
\epsfig{figure=uex.eps,height=6cm,width=8.4cm,angle=0}
\bigskip
\caption{Ratio of the potential energy to $N\, T$ of a one species cQGP as 
a function of the dimensionless coupling $\Gamma$.  The blue points are 
the MD simulations; the red curve is the fit in eq.~(\ref{uex_fit}).}
\label{uex}
\end{center}
\end{figure}

\section{Comparison with sQGP}

A qualitative mapping of the one species cQGP described above
into sQGP can be made with the
 by adjusting the three
basic scales described above, namely that of the length, 
the time and the mass. As we have 
repeatedly indicated above, all parameters of the model 
 are functions of the temperature
$T$ in the sQGP.  In this section
we provide heuristic arguments for what the parameters are using
a simple description of the sQGP at
 $T=1.5-3$ $T_c$. (It is used as a reference point
because more lattice data are
available for it than for $T$  close to $T_c$.)

As discussed above, the unit of length is $\lambda$ the minimum
of the potential. Since the repulsive part mocks up the quantum
repulsion at short distances, the effective inter-particle potential
is~\footnote{In so far, our arguments were classical and nonrelativistic, 
so both $\hbar$ and $1/c$ were set to zero.
 In this section, and this section only, we use the standard 
high energy units with $\hbar=1,c=1$. Hopefully,
 it will not create confusion for the reader. 
We will  keep $\hbar$ in the next formula, to emphasize the quantum origin of
the repulsive core.}

\be 
V_{\rm eff}= {\hbar^2\over 2mr^2}-\frac{C\,\alpha_s}r   
\label{veff}
\ee
with a minimum at  $r_0=\hbar^2/(mC\alpha_s)=\lambda$. $C$ is the pertinent
Casimir for quarks and gluons.
 
The quasiparticle mass, defined as its energy at zero momentum,
is at $T=(1.5-3)T_c$ about constant $m\approx 5 T_c$ 
both for quark and gluon quasiparticles~\cite{masses}.
We note that there are no direct 
lattice measurements of quasiparticle dispersion curves
below $1.5\,T_c$.  One can infer quark quasiparticle 
masses from fits to thermodynamical observables
and baryonic susceptibilities: it tells us that masses
grow toward $T_c$, for a recent discussion
see~\cite{Liao:2005pa}.  

For further discussion it is convenient to introduce
a dimensionless mass parameter $\tilde m=m/T$, which thus changes
from large values close to $Tc$ to about 3 at 1.5$T_c$. Clearly, 
its value determines how accurate is the nonrelativistic
approximation. For example, at $\tilde m=5$ the mean square
momentum $\sqrt{<p^2>}/T=4.8$ if calculated
relativistically, while the 
non-relativistic approximation gives the value smaller by a factor 0.8.
The corresponding ratio of the total density in non-relativistic
approximation is .7 of its relativistic value. So, it is not
an unreasonable approximation, but not very accurate. At $T=1.5T_c$
the value of this mass  is $\tilde m\approx 3$ only, and the accuracy of
it is even worse.
Therefore, we should calculate any momentum integrals
relativisitcally,
and only the results to be mapped into cQGP.

The color Casimir for gluons $C_g=3$ and quarks $C_q=4/3$
should be averaged over their respective weights. The number of
effective degrees of freedom in thermodynamical quantities
is such that roughly all three species -- $g,
\bar q ,q $ -- are equally represented. Furthermore,
the values of the coupling constant  $\alpha_s$ inferred
from measured static potentials at relevant distances is 
$\alpha_s\approx 0.5$. Withing the uncertainties, we will thus simply use
$<\alpha_s\,C>=1$ below. Thus

\be 
\lambda=r_0\approx \frac 1{3T}
\ee

The density of quasiparticles $n$ can be estimated as follows. 
Although quasiparticles are relatively heavy, the presence of bound 
states  compensates and increases the density of total quasiparticles
to about 0.8 of the entropy density per particle in the massless 
gas~\cite{THERMO}. Ignoring this 0.8 which refers to the bound states, 
we then have
\be 
n \approx (0.244\, T^3) \,(8+6\,N_f)\approx 6.3\, T^3 
\ee
where the first bracket is the usual
density of black body radiation photons, and the second bracket
is the number
of effective degrees of freedom due to 8 colored gluons and
2 quarks and antiquarks with 3  colors  and $N_f$ flavors
This corresponds to the following Wigner-Seitz radius

\be 
a_{WS}= \left(\frac 3{4\,\pi n}\right)^{1/3}
\approx {1\over 3 T}\approx \lambda
\ee 
The time units in MD are given by the plasma frequency, which is

\be 
\tau_0=\omega_p^{-1}= \left(\frac{4\pi n <\alpha_s\,C>}m\right)^{-1} 
\approx \frac 1{5.1\, T} 
\ee
where, for definiteness, we used $m=3T$.

In summary, the dimensionless cQGP results of the MD carried above can be 
qualitatively translated to the dimensionfull sQGP parameters
using respectively the length $\lambda$, time $\tau_0$ and mass $m$ units

\be
&&\lambda\approx \frac 1{3\,T}\nonumber\\
&&\tau_0\approx \frac 1{5.1\,T}\nonumber\\
&& m\approx 3\,T
\label{scales}
\ee
with  $\Gamma\approx 3$.

Indeed, the viscosity unit $\eta_0$ is then given by
the combination of units with the appropriate dimension

\be 
\eta_0= m/(\tau_0\,\lambda)=(3 T)(5.1 T)(3 T)\approx 46\, T^3  
\ee 
leading to the sQGP viscosity through

\be
\eta\rightarrow \eta\,\eta_0\approx (0.17)(46\,T^3)\approx 7.8\,T^3
\ee
In the sQGP, it is customary to give the viscosity
per entropy density, which  for ideal massless QGP is 
\be s_0={4\pi^2\over 90}\left(16+
\left({7\over 8}\right)\times 2 \times 2 \times 3 \times
N_f \right) T^3\approx 20 T^3\ee
for 3 flavors. Correcting a bit for more realistic lattice entropy, we
will use
$s\approx 23\,T^3$ around $T=1.5Tc$.
Thus, the dimensionless viscosity
$\eta\approx 0.17$ as measured in the one species cQGP above
at $\Gamma\approx 3$, can be translated to the dimensionfull 
viscosity in the sQGP by rescaling through $\eta_0$.
The viscosity per entropy ratio in the sQGP is

\be 
\frac {\eta}{s}\rightarrow \frac {\eta\,\eta_0}{s}\approx 0.34\,\,. 
\label{etas}
\ee
N=4 SYM results have suggested a universal lower bound for
$\eta/s\geq 1/4\pi$. Our analysis suggests that in the sQGP
this ratio is about 4 times the lower bound.

The diffusion constant translates to

\be
D\approx 0.161\rightarrow (0.161)(\lambda^2/\tau_0)\approx 0.1/T \,\,.
\ee
Similarly, the stress-tensor decorrelation time translates to

\be
\tau_\eta\approx 0.395\rightarrow (0.395)(\tau_0)\approx 0.08/T\,\,.
\ee
One should keep in mind, that these results are estimated for
$T\approx 1.5T_c$. As temperature gets closer to $T_c$, the masses
of quasiparticles grow further and the coupling gets larger, so
one finds even stronger coupled plasma. Unfortunately,
we do not yet have sufficiently precise lattice data to make
the mapping of the cQGP to the sQGP in this region more definite.

\section{Conclusions and prospects}

Quarks and gluons in the temperature range of 1-1.5 $T_c$ behave
as quasiparticles with masses  larger than
$m > 3\,T$ and a Casimir-weighted Coulomb strength of the
order of $<C\,\alpha_s>\approx 1$ which is strong. Lattice 
measured potentials in this temperature range give $\Delta U/T$ 
between 20 at $T_c$ and 5 at $1.2T_c$ indicating the dominance
of the potential energy over the thermal energy.

In this paper, we have suggested that the long-wavelength 
properties of the QGP in the 1-1.5 $T_c$ range may be modeled
by a classical and non-relativistic gas of massive quasi-particles
interacting via strong but classical color charges. The model we
coined cQGP has merits on its own and was analyzed numerically 
using molecular dynamics together with Wong's equations for 
classical color evolution.

Our results show the existence of
several phases ranging from a gas-like phase at weak coupling,
through a liquid-like phase at intermediate coupling, and 
finally a crystal-like phase at strong coupling with 
anti-ferromagnetic-like color ordering. The transition from
liquid to crystal is best seen in the density-density correlation
function or its static counterpart the structure function. 
At large Coulomb coupling the excess energy per particle
asymptotes the Madelung energy for a crystal.

We have used the numerical analysis to extract a number of 
transport coefficients including the quasiparticle diffusion
constant, the bulk viscosity and relaxation time. While
quantum mechanics is important in the sQGP in the formation
of the underlying quasiparticles (dispersion laws) and the
running of the coupling, we have suggested that the ensuing
quasiparticle interaction and dynamics is essentially Coulombic
and at strong coupling classical. For this, the cQGP
results are generic at intermediate and strong coupling. We
have given qualitative arguments for how to relate the cQGP
transport results to those of interest in the sQGP by 
identifying the pertinent length scales. Indeed, we have 
found that the sQGP corresponds to $\Gamma\approx 3$ which
is liquid-like and viscous with $\eta/s\approx 1/3T$.

In a series of sequels to follow, we will provide more insights to
the cQGP results presented above for the bulk, 2-particle 
correlations and transport properties.

\vskip 1.0cm

{\bf Acknowledgments.\,\,}

This work was partially supported by the US-DOE grants DE-FG02-88ER40388
and DE-FG03-97ER4014.

\end{narrowtext}
\end{document}